 \documentclass [12pt,a4paper      ]{article}
\usepackage{times}

\DeclareFontFamily{OT1}{times}{}
\DeclareFontShape {OT1}{times}{m }{n }{ <-> ptmr }{}
\DeclareFontShape {OT1}{times}{bx}{n }{ <-> ptmb }{}
\DeclareFontShape {OT1}{times}{m }{it}{ <-> ptmri}{}
\DeclareFontShape {OT1}{times}{bx}{it}{ <-> ptmbi}{}
\usepackage{amsmath}
\usepackage{amsfonts}
\usepackage{amssymb}
\usepackage{latexsym}
\usepackage{isridef} 
\numberwithin{equation}{section}               

\begin{document}

\title{\bf\vspace{-2.5cm} First-order quantum perturbation theory
                          and Colombeau generalized functions}

\author{
         {\bf Andre Gsponer}\\
         {\it Independent Scientific Research Institute}\\ 
         {\it Oxford, OX4 4YS, England}
       }

\date{ISRI-08-06.16 ~~ September 15, 2008}

\maketitle

\begin{abstract}

The electromagnetic scattering of a spin-0 charged particle off a fixed center is calculated in first-order quantum perturbation theory.  

This implies evaluating the square of a `Dirac delta-function,' an operation that is not defined in Schwartz distribution theory, and which in elementary text-books is dealt with according to `Fermi's golden rule.'  

In this paper these conventional calculations are carefully reviewed, and their crucial parts reformulated in a Colombeau algebra --- in which the product of distributions is mathematically well defined.

The conclusions are: (1) The Dirac delta-function insuring energy conservation in first order perturbation theory belongs to a particular subset of representatives of the Schwartz distribution defined by the Dirac measure.  These particular representatives have a well-defined square, and lead to a physically meaningful result in agreement with the data.  (2) A truly consistent mathematical interpretation of these representatives is provided by their redefinition as Colombeau generalized functions.  This implies that their square, and therefore the quantum mechanical rule leading from amplitudes to probabilities, is rigorously defined.

\end{abstract}

\section{Introduction}
\label{int:0}

Quantum theory and quantum field theory are plagued by fundamental conceptual problems which originate in part from the mathematical difficulties of properly interpreting and dealing with the `$\delta$-functions' which were first introduced in these theories by Dirac in 1927, i.e.,  \cite[p.\,624]{DIRAC1927-}, as a development of Lanczos's field theory \cite{LANCZ1926A}. 

In this paper we calculate in first-order time-dependent perturbation theory the cross-section of what is possibly the most simple interaction between two charged particles: The Rutherford scattering of a spin-0 particle of mass $m$ from a fixed Coulomb potential, i.e., the electric field of a nucleus of charge $Ze$ and mass $M \rightarrow \infty$.  The scattering center being fixed implies that the modulus of the incoming momentum $\vec{p}_i$ of the particle is equal to that of its outgoing momentum $\vec{p}_f$, i.e., $|\vec{p}_i|=|\vec{p}_f|$. Its final energy $E_f$ is thus equal to its initial energy $E_i$, and the only effect of the interaction is to change the direction of the incident particle.

All calculations are made in full detail, and at the most elementary level of quantum field theory.  

The objective is to clearly show how $\delta$-functions and their squares arise in that problem, and how these $\delta$- and $\delta^2$-functions are evaluated in conventional theory, as well as in Colombeau's theory of nonlinear generalized functions.

For an introduction to Colombeau generalized functions see \cite{COLOM1992-}, for a comprehensive textbook \cite{GROSS2001-}, and for a primer \cite{GSPON2006B} or \cite{GSPON2008A}.

\section{Definition and normalization of states}
\label{def:0}

The amplitude (i.e., the state, or wave-function) of a free spin-0 particle is defined as
\begin{equation}\label{def:1}
               \phi \DEF u \exp(i \vec{p}\cdot\vec{x}-i Et),
\end{equation}
where $u \in \mathbb{C}$.  The probability current density associated to this amplitude is the four-vector
\begin{equation}\label{def:2}
           J_\mu = i \Bigl( \phi^*(\partial_\mu \phi) -
                            (\partial_\mu \phi^*)\phi   \Bigr),
\end{equation}
so that the probability density, given by the time component of that four-vector, is
\begin{equation}\label{def:3}
  \frac{d\mathcal{P}}{d^3x} = i \Bigl( \phi^*(\partial_t \phi) -
                            (\partial_t \phi^*)\phi   \Bigr).
\end{equation}

   The state $\phi$ is assumed to be an element of a Hilbert space in which the Lorentz invariant scalar product is defined as
\begin{equation}\label{def:4}
  \BRA \phi_1 \| \phi_2 \KET \DEF i \iiint_\mathcal{V} 
  \Bigl( \phi_1^*(\partial_t \phi_2) - (\partial_t \phi_1^*)\phi_2 \Bigr).
\end{equation}
The condition normalizing the amplitude \eqref{def:1} can therefore be written
\begin{equation}\label{def:5}
  \mathcal{P} = \iiint_\mathcal{V} d\mathcal{P} = \BRA \phi \| \phi \KET = 1,
\end{equation}
where $\mathcal{V} \subset \mathbb{R}^3$ is a large `box' of volume $V$ in which the scattering experiment is supposed to take place.  Substituting $\phi$ from \eqref{def:1} this yields
\begin{equation}\label{def:6}
  \mathcal{P} = i \iiint_\mathcal{V} d^3x~ u^*u(-iE-iE) = 2u^*u EV  = 1,
\end{equation}
which implies that the normalized amplitude is
\begin{equation}\label{def:7}
               \phi = \frac{1}{\sqrt{2EV}}
               \exp(i \vec{p}\cdot\vec{x}-i Et + i\varphi_0),
\end{equation}
where the constant phase $\varphi_0$ will be set to zero in the following.

\section{Calculation of the transition amplitude}
\label{amp:0}

   In first order perturbation theory the transition matrix element from a state $\phi_i$ to a state $\phi_f$ is given by
\begin{equation}\label{amp:1}
         S_{fi} = 2 e \iiint_{\mathbb{R}^3} d^3x
                      \int_{\mathbb{R}} dt~\phi_f^* \mathbf{A} \phi_i,
\end{equation}
where $\mathbf{A}$ is the operator corresponding to the perturbation.  In our case $\mathbf{A}$ is simply the Coulomb potential of the nucleus, i.e., in Heaviside units,
\begin{equation}\label{amp:2}
         \mathbf{A} = \frac{Ze}{4\pi|\vec{x}\,|}.
\end{equation}
  Thus, substituting the normalized initial and final state amplitudes,
\begin{equation}\label{amp:3}
     S_{fi} = \frac{Ze^2}{4\pi V} \frac{1}{\sqrt{E_f E_i}}
    \iiint_{\mathbb{R}^3} \frac{d^3x}{|\vec{x}\,|}
                          \exp\Bigl(i (\vec{p}_i-\vec{p}_f)\cdot\vec{x}\Bigr)
    \int_{-\infty}^{+\infty} dt~ \exp\Bigl(i (E_f-E_i)t \Bigr).
\end{equation}
The spatial integral is the Fourier transform of the Coulomb potential,\footnote{Our conventions are such that the Fourier transform is $\FOU{f}(p) = \int_{-\infty}^{+\infty} \exp(ipx)~f(x)~dx$ and its inverse $f(x) = (1/2\pi)\int_{-\infty}^{+\infty} \exp(-ixp)~\FOU{f}(p)~dp$.} well known to be
\begin{equation}\label{amp:4}
      \iiint_{\mathbb{R}^3} \frac{d^3x}{|\vec{x}\,|}
      \exp\Bigl(i (\vec{p}_i-\vec{p}_f)\cdot\vec{x}\Bigr)
    = \frac{4 \pi}{|\vec{q}\,|^2},
\end{equation}
where $\vec{q} = \vec{p}_f - \vec{p}_i$ is the transferred momentum.  The integral over the time coordinate is not well defined because its integrand is non zero at infinity.  Nevertheless, in conventional calculations, it is interpreted as yielding the $\delta$-function\footnote{What is meant here by `$\delta$-function' or `Dirac function' is the loosely-defined object known under these names by the physicists, not the mathematically well defined `Dirac measure' or `$\delta$-distribution' of Schwartz distribution theory.}
\begin{equation}\label{amp:5}
     \int_{-\infty}^{+\infty} dt~\exp\Bigl(i (E_f-E_i)t \Bigr)
     = 2\pi \, \delta(E_f-E_i),
\end{equation}
which expresses energy conservation between the initial and final states in a static potential.  Thus, with $q = |\vec{q}\,|$, the transition amplitude is
\begin{equation}\label{amp:6}
     S_{fi} = 2\pi \frac{Ze^2}{V} \frac{1}{\sqrt{E_f E_i}}
                   \frac{1}{q^2} \delta(E_f-E_i).
\end{equation}
\begin{remark}[Conventional interpretation of the $\delta$-function]  
\label{rema:1}
Eq.~\eqref{amp:5} is mathematically ill-defined because $\exp(ix)$ is undefined at $x=\pm \infty$.  The conventional interpretation in quantum theory assumes that there is an implicit cut-off such that 
\begin{equation}\label{amp:7}
   \delta(x) \DEF
   \frac{1}{2\pi} \lim_{T \rightarrow\infty} \int_{-T/2}^{+T/2} dp~\exp(\pm ixp)
 = \frac{1}{2\pi} \lim_{T \rightarrow\infty} \int_{-T/2}^{+T/2} dp~\cos(xp),
\end{equation}
is real-valued and is even in the variable $x$.
\end{remark}

\section{Calculation of the transition probability}
\label{pro:0}

    The transition probability $dW_{fi}$ is obtained by multiplying the modulus squared of the transition amplitude by the phase-space element, that is by the number of final states in momentum interval $d^3p_f$, which is equal to $V d^3p_f / (2\pi)^3$.  Therefore
\begin{equation}\label{pro:1}
                dW_{fi} =  |S_{fi}|^2 \frac{V d^3p_f}{(2\pi)^3}.
\end{equation}
After using \eqref{amp:6} and making the substitution $e^2 \rightarrow 4\pi \alpha$, where $\alpha=1/137$ is the electromagnetic coupling constant, this transition probability becomes
\begin{equation}\label{pro:2}
   dW_{fi} =  8\pi \frac{Z^2\alpha^2}{E_iV} \frac{1}{q^4}
                   \frac{d^3p_f}{E_f} |\delta|^2(E_f-E_i).
\end{equation}
in which the modulus squared of the $\delta$-function arises.

  The time-integrated total cross section $\Sigma$ is defined by the phase-space integrated transition probability divided by the incoming flux, that is divided by $v_i=|\vec{v}_i\,|/V$ where $\vec{v}_i = \vec{p}_i/E_i$ is the incident velocity.  Thus
\begin{equation}\label{pro:3}
                \Sigma \DEF  \iiint_{\mathbb{R}^3} dW_{fi} \frac{V}{v_i},
\end{equation}
and therefore
\begin{equation}\label{pro:4}
    \Sigma = 8\pi Z^2\alpha^2 \iiint_{\mathbb{R}^3}
                   \frac{1}{E_iv_i} \frac{1}{q^4}
                   \frac{d^3p_f}{E_f} |\delta|^2(E_f-E_i).
\end{equation}
Then, with $\vec{v}_i E_i = \vec{p}_i$, and $d^3p_f = d\Omega \, p_f^2 dp_f$,
\begin{equation}\label{pro:5}
  \frac{d\Sigma}{d\Omega} = 8\pi Z^2\alpha^2
                            \int_0^{+\infty}\frac{p_f}{p_i} \frac{1}{q^4}
                            |\delta|^2(E_f-E_i)  \frac{p_f dp_f}{E_f}.
\end{equation}
Finally, since by definition $E^2 - p^2 = m^2$ we have $p_f dp_f = E_f dE_f$, and thus
\begin{equation}\label{pro:6}
    \frac{d\Sigma}{d\Omega} = 8\pi Z^2\alpha^2
         \int_{-\infty}^{+\infty} \frac{p_f}{p_i}
         \frac{1}{q^4} |\delta|^2(E_f-E_i) dE_f,
\end{equation}
where, since $\delta(E_f-E_i) \neq 0$ only for $E_f \approx E_i$, the integration range was extended from $[0,+\infty]$ to $[-\infty,+\infty]$  for convenience.

\section{Calculation of the differential cross section}
\label{dif:0}

Equation \eqref{pro:6} contains $|\delta|^2$, which like $\delta^2$ is mathematically undefined if $\delta$ is interpreted as a Schwartz distribution.  In standard quantum theory this fact is generally ignored and $\delta$ is assumed to be a real valued function defined by a regularization consistent with Remark \ref{rema:1}.  The conventional method for calculating \eqref{pro:6} can then be introduced in several ways.  A purely heuristic approach consists of blindly generalizing the usual sifting property of the $\delta$-distribution, i.e.,
\begin{equation}\label{dif:1}
    \int_{-\infty}^{+\infty} f(E_f-E_i) \, \delta(E_f-E_i) \, dE_f = f(0),
\end{equation}
to the $\delta^2$-function by writing
\begin{align}
\nonumber
    &\int_{-\infty}^{+\infty} f(E_f-E_i) \, \delta^2(E_f-E_i) \, dE_f\\
\label{dif:2}
  = &\int_{-\infty}^{+\infty}
     \Bigl( f(E_f-E_i) \, \delta(E_f-E_i) \Bigr) \delta(E_f-E_i) \, dE_f
  = f(0) \, \delta(0).
\end{align}
With $\delta(E)$ defined according to \eqref{amp:5}, the diverging quantity $\delta(0)$ is then estimated by introducing a symmetric cut-off to regularize the integral as in \eqref{amp:7}, i.e.,
\begin{equation}\label{dif:3}
    \delta(E) = \frac{1}{2\pi} \int_{-\infty}^{+\infty} dt~\exp(i Et ) 
              = \frac{1}{2\pi} \lim_{T \rightarrow \infty}
                \int_{-T/2}^{+T/2} dt~\exp(i Et ),
\end{equation}
so that
\begin{equation}\label{dif:4}
    \delta(0) = \lim_{T \rightarrow \infty} \frac{T}{2\pi}.
\end{equation}
Applying \eqref{dif:2} and \eqref{dif:4} to \eqref{pro:6}, and dividing both sides by $T$, the time-integrated cross-section becomes the differential scattering cross section
\begin{equation}\label{dif:5}
     \frac{d\sigma}{d\Omega} 
     \DEF  \lim_{T \rightarrow \infty}  \frac{d\Sigma}{T d\Omega}
      = \frac{4 Z^2\alpha^2}{q^4},
\end{equation}
which is known as the Rutherford scattering formula for a spin-0 projectile.  This result is very well confirmed by experiment as the first term of a perturbation series.\footnote{If the projectile were a spin-$1/2$ particle Eq.\,\eqref{dif:5} would be multiplied by a correction factor that would reduce to $1$ in the non-relativistic limit.}  For this reason, any regularization of the $\delta$-function which like \eqref{dif:3} yields \eqref{dif:4} for $\delta(0)$ is a suitable representation of the $\delta$-function for this quantum mechanical problem.  

  It should also be remarked that the physically measurable quantity, the cross section \eqref{dif:5}, is a rate `averaged' over a very long time $T$, not an instantaneous rate in a infinitesimal interval $dT$.  Actually, the time $T$ is the `duration' of the measurement.  Its product with the the volume $V$ of the `box' used to normalize the amplitude $\phi$ defines a space-time volume $VT$ such that the integration over infinite space-time in \eqref{amp:1} is actually restricted to that finite hyper-volume.

This heuristic derivation can be improved in various ways, see for example \cite[p.\,101]{BJORK1964-}.  Moreover, the same method can be applied to other first order perturbation problems for which a general formula, called `Fermi's golden rule,' can be derived, see for example \cite[p.\,332]{SAKUR1994-}.  Finally, the method can be further generalized to four-dimensions and thus to invariant perturbation theory, e.g.,  \cite[p.\,112]{BJORK1964-}, \cite[p.\,164]{JAUCH1955-}.


\section{Justification of the conventional method}
\label{jus:0}

In this section we show that the $\delta$-function regularized according to the standard prescription, given in Remark \ref{rema:1}, is consistent with Eqs.\,\eqref{dif:1} and \eqref{dif:2}. 

Let us rewrite \eqref{dif:2} in the form
\begin{align}\label{jus:1}
                   \int_{-\infty}^{+\infty} f(x) \, \delta^2(x) \, dx
       = \delta(0) \int_{-\infty}^{+\infty} f(x) \, \delta(x) \, dx,
\end{align}
where, replacing $T/2$ by $1/\epsilon$ with $0 < \epsilon < 1$ and making an elementary integration,  the $\delta$-function defined by \eqref{dif:3} corresponds to
\begin{equation}\label{jus:2}
    \delta(x) = \lim_{\epsilon \rightarrow 0} \frac{1}{2\pi}
                \int_{-1/\epsilon}^{+1/\epsilon} \exp(-i p x) ~dp
              = \lim_{\epsilon \rightarrow 0} \frac{1}{\pi}
                \frac{\sin(x/\epsilon)}{x}.
\end{equation}
Because, \cite[p.\,405]{GRADS1965-},
\begin{align}\label{jus:3}
     \int_{-\infty}^{+\infty} 
     \frac{1}{\pi} \frac{\sin(x/\epsilon)}{x} ~dx = 1,
\end{align}
the sequence
\begin{equation} \label{jus:4}
     \rho_\epsilon \DEF \frac{1}{\epsilon} \rho\Bigl(\frac{x}{\epsilon}\bigr),
     \qquad \text{where} \qquad
     \rho(x) = \frac{1}{\pi} \frac{\sin(x)}{x},
\end{equation}
is normalized to $1$.  However, since $\rho_\epsilon$ has non-compact support, we have to carefully verify under which conditions $\rho_\epsilon$ is a $\delta$-sequence.  Thus we take $f \in \mathcal{D}$, i.e., a smooth function with compact support, and calculate
\begin{align}
\nonumber
    \int_{-\infty}^{+\infty} f(x) \, \delta(x) \, dx
    &= \lim_{\epsilon \rightarrow 0}
                \frac{1}{\pi} \int_{-\infty}^{+\infty} f(x)
                \frac{\sin(x/\epsilon)}{x} ~dx\\
\label{jus:5}
    &= \lim_{\epsilon \rightarrow 0}
                \frac{1}{\pi} \int_{-\infty}^{+\infty}
                 f(\epsilon z) \frac{\sin(z)}{z} ~dz.
\end{align}
Then, if we try to develop $f(\epsilon z)$ in a Taylor series, we are immediately confronted with the difficulty that the moments
\begin{align} \label{jus:6}
     \int_{-\infty}^{+\infty} 
     \frac{1}{\pi} \frac{\sin(z)}{z} z^{2n}~dz = \infty,
     \qquad \forall n \geq 1.
\end{align}
We therefore try another standard method \cite[p.\,29]{SCHUC1991-}, which consists of splitting the integral into three pieces, i.e., 
\begin{align}
\label{jus:7}
    \int_{-\infty}^{+\infty} f(x) \, \delta(x) \, dx
    &= \lim_{\epsilon \rightarrow 0}
                \frac{1}{\pi} \Bigl(
                \int_{-\infty}^{-\sqrt{\epsilon}}
              + \int_{-\sqrt{\epsilon}}^{+\sqrt{\epsilon}}
              + \int_{+\sqrt{\epsilon}}^{+\infty} \Bigr)
                f(x) \frac{\sin(x/\epsilon)}{x} ~dx.
\end{align}
The first and the third pieces tend to zero because $f \in \mathcal{D}$ implies that there is a bound $B $ such that
\begin{align}
\label{jus:8}
        \Bigl| \lim_{\epsilon \rightarrow 0}
                \frac{1}{\pi}
                \int_{\pm \infty}^{\pm \sqrt{\epsilon}}
                 f(x) \frac{\sin(x/\epsilon)}{x} ~dx ~\Bigr|
     < \Bigl| B \lim_{\epsilon \rightarrow 0}
                \frac{1}{\pi}
                \int_{\pm \infty}^{\pm 1/\sqrt{\epsilon}}
                \frac{\sin(z)}{z} ~dz ~\Bigr|
            = 0.
\end{align}
The second piece can be computed using the mean value theorem because $\sin(z)/z$ is even.  Thus, with $\xi \in [-\sqrt{\epsilon}, +\sqrt{\epsilon}\,]$,
\begin{align}
\label{jus:9}
         \lim_{\epsilon \rightarrow 0}
                \frac{1}{\pi}
                \int_{-\sqrt{\epsilon}}^{+\sqrt{\epsilon}}
                 f(x) \frac{\sin(x/\epsilon)}{x} ~dx
             =  \lim_{\epsilon \rightarrow 0} f(\xi)
                \frac{1}{\pi}
                \int_{-1/\sqrt{\epsilon}}^{+1/\sqrt{\epsilon}}
                \frac{\sin(z)}{z} ~dz
             =  f(0).
\end{align}
Therefore, provided $f \in \mathcal{D}$, $\delta$ defined by \eqref{jus:2} has the sifting property of the Dirac distribution, i.e., Eq.\,\eqref{dif:1}, thanks to the normalization \eqref{jus:3}.  Similarly, since \cite[p.\,446]{GRADS1965-}
\begin{align}\label{jus:10}
     \int_{-\infty}^{+\infty} \rho_\epsilon^2(x) ~dx
   = \int_{-\infty}^{+\infty} 
     \frac{1}{\pi^2} \frac{\sin^2(x/\epsilon)}{x^2} ~dx
   = \frac{1}{\pi\epsilon},
\end{align}
the same proof works for the square of $\delta_\epsilon$.  Indeed,
\begin{align}
\label{jus:11}
    \int_{-\infty}^{+\infty} f(x) \, \delta^2(x) \, dx
    &= \lim_{\epsilon \rightarrow 0}
                \frac{1}{\pi^2} \int_{-\infty}^{+\infty} f(x)
                \frac{\sin^2(x/\epsilon)}{x^2} ~dx\
     = \lim_{\epsilon \rightarrow 0}  \frac{1}{\pi\epsilon} f(0),
\end{align}
implies that the sequence $\pi\epsilon \rho_\epsilon^2(x)$ also defines a $\delta$-function.  Moreover, $\delta(0)= \lim_{\epsilon \rightarrow 0} 1/(\pi\epsilon)$, as shown by \eqref{jus:2}.  The right-hand side of \eqref{jus:11} is thus equal to that of \eqref{jus:1}, which is therefore proved.

   Consequently, the conventional calculations leading to the Rutherford formula \eqref{dif:5} are correct.  However, these calculations are entirely based on the special properties of the particular representative \eqref{jus:2} of the $\delta$-function.  More precisely, for $f \in \mathcal{D}$, the proof of \eqref{jus:1} requires that the $\delta$-sequence $\rho_\epsilon$ must be such that $\rho$ has the two properties
\begin{align}
\label{jus:12}
     & \int_{-\infty}^{+\infty} \rho(z) ~dz = 1,\\
\intertext{and}
\label{jus:13}
     & \int_{-\infty}^{+\infty}  \rho^2(z) \, dz = \rho(0),
\end{align}
and the Rutherford formula is obtained if moreover
\begin{align}\label{jus:14}
                 \rho(0) = \frac{1}{\pi}.
\end{align}

In conclusion, the $\delta$-function defined by \eqref{amp:5} and regularized according to \eqref{amp:7} is a special representative of the Schwartz $\delta$-distribution having the additional properties \eqref{jus:13} and  \eqref{jus:14}.  The mathematical object denoted `$\delta$' in Eq.\,\eqref{pro:6}, and named `$\delta$-function' by the physicists, is therefore \emph{not} a Schwartz distribution in the sense that its representatives $\rho_\epsilon$ are not \emph{any} element of the equivalence class of representatives corresponding to the functional $\delta(f)$ defined by Eq.\,\eqref{dif:1}.

\section{Remarks on the conventional method}
\label{rem:0}

\begin{enumerate}

\item The $\delta$-function \eqref{jus:2} can be rewritten as
\begin{equation}\label{rem:1}
  \delta(x) = \lim_{\epsilon \rightarrow 0} \rho_\epsilon(x)
            = \lim_{\epsilon \rightarrow 0}  \frac{1}{2\pi}
                 \int_{-\infty}^{+\infty}
                 \exp(-i px ) \FOU{\rho}(\epsilon p) ~dp
\end{equation}
where instead of being implemented by finite bounds in the integration range, the `sharp' regularizing cut-off is implemented by the Fourier transform of $\rho(x)$, i.e., by the piece-wise continuous function $\FOU{\rho}(z) = 1$ for $|z| \leq 1$, and $\FOU{\rho}(z) = 0$  for $|z| > 1$.

  Similarly, the conditions \eqref{jus:12} and \eqref{jus:14} on $\rho$, i.e.,
\begin{equation}\label{rem:2}
     \int_{-\infty}^{+\infty} \rho(z) ~dz = 1,
 \qquad  \text{and} \qquad
     \rho(0) = \frac{1}{\pi},
\end{equation}
can be implemented on its Fourier transform $\FOU{\rho}$, i.e.,
\begin{equation}\label{rem:3} 
    \FOU{\rho}(0) = 1,
 \qquad  \text{and} \qquad
     \int_{-\infty}^{+\infty} \FOU{\rho}(z) ~dz = 2.
\end{equation}

Thus, any cut-off function $\FOU{\rho} \in \mathbb{R}$ satisfying \eqref{rem:3} as well as \eqref{jus:13} would yield a regularization of \eqref{amp:5} leading to the Rutherford formula, and more generally to any formula derived by means of Fermi's golden rule.
 
For example, the exponential damping factor $\FOU{\rho}^E = \exp(-|z|)$ satisfies the conditions \eqref{rem:3}.  It provides a `smooth' cut-off such that
\begin{equation}\label{rem:4}
    \rho_\epsilon^E(x)  = \frac{1}{2\pi}
                \int_{-\infty}^{+\infty} dp~\exp(-i px )
                \exp( -\epsilon |p|)
 = \frac{1}{\pi \epsilon} \frac{\epsilon}{x^2 + \epsilon^2},
\end{equation}
which is a well-know representative of the $\delta$-function having the property $\delta(0)_\epsilon^E=1/(\pi\epsilon)$.  Unfortunately, $\rho_\epsilon^E$ does not satisfy  \eqref{jus:13}.   As a second example, if the regularization is made by means of the Gaussian damping factor $\FOU{\rho}^G = \exp(-z^2/4)$, one gets another well-know representative of the $\delta$-function, i.e.,
\begin{equation}\label{rem:5}
    \rho_\epsilon^G(x)  = \frac{1}{2\pi}
                \int_{-\infty}^{+\infty} dp~\exp(-i px )
                \exp\Bigl( -\frac{\epsilon^2 p^2}{4} \Bigr)
 = \frac{1}{\sqrt{\pi} \epsilon} \exp\Bigl( -\frac{x^2}{\epsilon} \Bigr),
\end{equation}
which however does neither satisfy \eqref{jus:3} nor \eqref{jus:14} because $\delta(0)_\epsilon^G=1/\sqrt{\pi} \epsilon$.  It is therefore not easy to find a simple example of a smooth damping function satisfying all three conditions (\ref{jus:12}, \ref{jus:13}, and \ref{jus:14}).

\item In the notation of Sec.~\ref{dif:0}, where $1/\epsilon$ corresponds to $T/2$, the conditions \eqref{rem:3} read
\begin{equation}\label{rem:6}
    \FOU{\rho}(0) = 1,
 \qquad  \text{and} \qquad
     \int_{-\infty}^{+\infty} \FOU{\rho}\Bigl( \frac{2t}{T} \Bigr) ~dt = T.
\end{equation}
Therefore, the physical interpretation of the second condition on $\FOU{\rho}$, which in the form \eqref{rem:2} corresponds to the point value $\rho(0)=1/\pi$, is simply that its time integral is equal to $T$, the effective time over which the scattering process is averaged (or measured).  

\item To prove \eqref{jus:1} we assumed that $f \in \mathcal{D}$, i.e., that $f$  was smooth and had compact support.  However, only the facts that $f \in \mathcal{C}^\infty$ and $f < B$ where actually used in the proof.  Let us verify that the function $f(E)$ implied by \eqref{pro:6} satisfies these conditions.  We have
\begin{equation}\label{rem:7}
    f(E_f,E_i) = \frac{p_f}{p_i} \frac{1}{q^4},
\end{equation}
where $p = \sqrt{E^2-m^2}$ and where, with $\theta$ the angle between $\vec{p}_f$ and $\vec{p}_i$,
\begin{equation}\label{rem:8}
    q^2(E_f,E_i) = |\vec{p}_f - \vec{p}_i|^2 
                 = p_f^2 + p_i^2 - 2 p_f p_i \cos \theta.
\end{equation}
Thus, for fixed $E_i \neq m$, i.e., $p_i \neq 0$, the function 
\begin{equation}\label{rem:9}
    f(E_f,E_i) = \frac{p_f}{p_i}
                 \frac{1}{(p_f^2 + p_i^2 - 2 p_f p_i \cos \theta)^2},
\end{equation}
is $\mathcal{C}^\infty$ and bounded, unless $E_f=E_i$ and $\theta =0$.  Consequently, excluding the latter case (which arises in the limit where there is no interaction) $f$ does indeed satisfy the conditions under which \eqref{jus:1} is valid.

\end{enumerate}

\section{Rigorous method}
\label{rig:0}

There are several way of treating the physical problem addressed in this paper in a Colombeau algebra.  The most comprehensive approach would be to define all objects as elements of a Colombeau algebra right from the beginning, as is done for example in \cite{COLOM2008-}.  This would provide a formulation having all the advantages of the `wave-packet' descriptions of scattering processes \cite[p.\,384]{SAKUR1994-}.  However, for the present problem, it is enough to suppose that the damping function $\FOU{\rho}$ employed to regularize the $\delta$-function \eqref{rem:1} is a suitable damper insuring that the $\delta$-sequence $\rho_\epsilon$ is an element of a Colombeau algebra.\footnote{For the definitions of suitable mollifiers and dampers in a Colombeau algebra, and for a discussion of their basic properties, see. e.g., \cite[Sec.\,1.4]{COLOM2008-}.}  This is sufficient for the modulus squared of the $\delta$-function arising when squaring $S_{fi}$ given by \eqref{amp:6} --- e.g., when calculating the probability \eqref{pro:1} --- to be mathematically well defined and meaningful. 

   The conditions for $\FOU{\rho}$ to be a suitable damper are simply that
\begin{equation} \label{rig:1}
     \FOU{\rho} \in \mathcal{D}(\mathbb{R}),
     \quad \text{and} \quad
      \FOU{\rho}(0) \equiv 1,
\end{equation}
which by Fourier transformation are equivalent to the well-known Colombeau constraints on the moments of $\rho$, i.e.,\footnote{Note that if the damper $\FOU{\rho} \in \mathbb{R}$ the mollifier $\rho$ may nevertheless be complex.}
\begin{equation} \label{rig:2}
    \rho \in \mathcal{S}(\mathbb{C}),
    \quad
    \int dz~\rho(z) = 1,
    \quad \text{and} \quad
    \int dz~z^n\rho(z) = 0, \quad \forall n=1,...,q,
\end{equation}
where $q \in \mathbb{N}$ is as large as we please.

   To these conditions we need to add the special constraints (\ref{jus:13}--\ref{jus:14}),  i.e., 
\begin{equation}\label{rig:3}
     \int |\rho|^2(z) \, dz  =  \rho(0),
 \qquad  \text{and} \qquad
     \rho(0) = \rho^*(0) = \frac{1}{\pi},
\end{equation}
which insure that the properties of the damper $\FOU{\rho}$  acting as a `smooth cut-off' are equivalent to thoses of the `sharp cut-off' introduced in \eqref{dif:3} with regards to the present problem, i.e., to obtain the Rutherford formula.  Note that, since $\FOU{\rho} \in \mathbb{R}$, the second condition on the right of \eqref{rig:3} is consistent because by Fourier transform $\rho(0) = (2\pi)^{-1}\int \FOU{\rho}(p) \,dp$ is necessarily real, even though $\rho(x)$ may be complex for $x\neq 0$.

   To find out the full consequencies of working in the Colombeau setting we reconsider \eqref{jus:1} in which $\delta(x) \in \mathbb{C}$ defined by the representative sequence $\rho_\epsilon$ is now subject to the Colombeau constraints \eqref{rig:2} and to the physical constraints \eqref{rig:3}.  That is, we study
\begin{align}\label{rig:4}
         \lim_{\epsilon \rightarrow 0}
         \int f(x) \, |\rho_\epsilon|^2(x) \, dx
       = \lim_{\epsilon \rightarrow 0}
         \rho_\epsilon(0) \int f(x) \, \rho_\epsilon(x) \, dx,
\end{align}
where for the sake of generality we suppose that $f \in \mathcal{O}_{\text{M}}(\mathbb{R},\mathbb{C})$, i.e., $f(x) \in \mathcal{C}^\infty$ and each of its derivatives do not grow faster than a power of $x$ at infinity in $\mathbb{R}$.  Thus, making the change of variable $x=\epsilon z$, we want to find out under which conditions we have
\begin{align}\label{rig:5}
          \lim_{\epsilon \rightarrow 0}
          \int f(\epsilon z) \, |\rho|^2(z) \, dz
        = \lim_{\epsilon \rightarrow 0}
          \rho(0) \int f(\epsilon z) \, \rho(z) \, dz.
\end{align}

  We begin with the right-hand side of \eqref{rig:5} and use Taylor's theorem with remainder.  We obtain
\begin{align}
\nonumber
    &\lim_{\epsilon \rightarrow 0}
     \rho(0) \int_{-\infty}^{+\infty} \rho(z) f(\epsilon z) ~dz\\
\nonumber
  =\, &\rho(0) \lim_{\epsilon \rightarrow 0} \int_{-\infty}^{+\infty} \rho(z)
  \Bigl(f(0) + \epsilon zf'(0) + \frac{(\epsilon z)^2}{2!}f''(0) + ...\Bigr)dz\\
\nonumber
  =\, &\rho(0)f(0)
     + \lim_{\epsilon \rightarrow 0} \rho(0) \int_{-\infty}^{+\infty} \rho(z)
            \frac{(\epsilon z)^{q+1}}{(q+1)!}f^{(q+1)}(\vartheta z)~dz\\
\label{rig:6}
  =\, &\rho(0) f(0) + \lim_{\epsilon \rightarrow 0} \OOO(\epsilon^{q+1}), 
\end{align}
where we used the Colombeau constraints \eqref{rig:2} implying that all moments of order $n$ with $1 < n < q+1$ are identically zero, and where $\vartheta \in ]0,1[$ in the remainder.  Then we observed that since $\rho \in \mathcal{S}$ and $f^{(q+1)} \in \mathcal{O}_{\text{M}}$ their product with $z^{q+1}$ is still in $\mathcal{S}$, and therefore the remainder is bounded so that \eqref{rig:6} is equal to $f(0) + \OOO(\epsilon^{q+1})$.

   Turning to the left-hand side of  \eqref{rig:5} and making again a Taylor development we get,
\begin{align}
\nonumber
   &\lim_{\epsilon \rightarrow 0} \int_{-\infty}^{+\infty}
    |\rho|^2(z) f(\epsilon z) ~dz\\
\nonumber
 =\, &\lim_{\epsilon \rightarrow 0}
    \int_{-\infty}^{+\infty} |\rho|^2(z) 
    \Bigl(f(0) +\epsilon zf'(0) +\frac{(\epsilon z)^2}{2!}f''(0) + ...\Bigr)dz\\
\nonumber
 =\, &f(0) \int_{-\infty}^{+\infty} |\rho|^2(z)~dz 
    + \lim_{\epsilon \rightarrow 0} \Bigl(
      \epsilon f'(0)\int_{-\infty}^{+\infty} z|\rho|^2(z)~dz
        + ... +\OOO(\epsilon^{q+1}) \Bigr)\\
\label{rig:7}
 =\, &\rho(0)f(0) 
    + \lim_{\epsilon \rightarrow 0} \Bigl(
      \epsilon f'(0)\int_{-\infty}^{+\infty} z|\rho|^2(z)~dz
        + ... +\OOO(\epsilon^{q+1}) \Bigr),
\end{align}
where we used \eqref{rig:3} to get the last line, and where we did not write the remainder explicitly because its form is similar to that in \eqref{rig:6} with $\rho$ replaced by $|\rho|^2$, so that it is again $\OOO(\epsilon^{q+1})$.  Equation \eqref{rig:7} cannot be further reduced unless specific constraints are put on the moment of $|\rho|^2$.  Due to $\rho \in \mathcal{S}$, the only thing that can be said is that these moments are finite.

  Comparing \eqref{rig:6} and \eqref{rig:7} we see that equation \eqref{rig:4} can in principle be obtained at two different orders of approximation:

\begin{enumerate}

\item \emph{Eq.}~\eqref{rig:4} \emph{is an equality $\OOO(1)$ as $\epsilon \rightarrow 0$}.

This case corresponds to the conventional perturbation theory calculations in which the results are limits such as \eqref{dif:5} where it does not matter how fast the limit is approached.  This is a suitable method if the result is not to be used in a subsequent calculation in which it could be multiplied by a quantity which might be diverging so that the $\OOO(1)$ remainder can be safely ignored relative to the leading term which is $\OOO(1/\epsilon)$.  If that is the case, no special constraint on the moments of $|\rho|^2$ with $n \neq 0$ is required, and the only constraints on the damper $\FOU{\rho}$ are those implied by \eqref{rig:1} and \eqref{rig:3}.  Expressing them as conditions on its inverse Fourier transform $\rho$, they are
\begin{align}
 \nonumber
    \rho(z) \in \mathcal{S}(\mathbb{C}),
    \quad
   &\int dz~\rho(z) = 1,
    \quad
    \int dz~z^n\rho(z) = 0,\\
\label{rig:8}
    \rho(0) = \frac{1}{\pi},
    \quad
   &\int dz~ |\rho|^2(z)  = \rho(0).
\end{align}

\item \emph{Eq.}~\eqref{rig:4} \emph{is an equality $\OOO(\epsilon^q)$ for all $q \in \mathbb{N}$ provided $\epsilon \in ]0,1[$}.

This corresponds to the case where \eqref{rig:4} would be an equality in the Colombeau algebra, which implies that the difference between the two sides of  \eqref{rig:4} must be  $\OOO(\epsilon^q)$.  Referring to \eqref{rig:7} this requires that all the moments of $|\rho|^2$ with $n = 1,2,...,q$ are zero.  Therefore, the full set of conditions on $\rho$ is
\begin{align}
\nonumber
    \rho(z) \in \mathcal{S}(\mathbb{C}),
    \quad
    \int dz~\rho(z) = 1,
    \quad
   &\int dz~z^n\rho(z) = 0,\\
\label{rig:9}
    \rho(0) = \frac{1}{\pi},
    \quad
    \int dz~ |\rho|^2(z)  = \rho(0),
    \quad
   &\int dz~z^n|\rho|^2(z) = 0.
\end{align}
However, as $|\rho|^2 \in \mathbb{R}$, it is immediately seen that the last condition on the second line of \eqref{rig:9} cannot be satisfied when $n$ is even, unless $\rho$ is identically zero.  Consequently, the option ``Eq.~\eqref{rig:4} is an equality $\OOO(\epsilon^q)$'' is \emph{not} feasible.

\end{enumerate}

\section{Discussion}
\label{dis:0}

There are three major mathematical problems in calculating the cross-section of the scattering process considered in this paper:
\begin{enumerate}

\item The interpretation of the time-integral in Eq.\,\eqref{amp:3} as a Dirac $\delta$-function insuring energy-conservation in the transition amplitude \eqref{amp:6};

\item The interpretation of the square of that $\delta$-function in the transition probability \eqref{pro:2}, which derives from the quantum-mechanical rule \eqref{pro:1} associating a probability to the modulus squared of the amplitude of the process;

\item The integration of that $\delta^2$-function in the differential cross-section \eqref{pro:6}.

\end{enumerate}

The conventional solution to the first problem is to introduce a sharp cut-off to regularize the time-integral in Eq.\,\eqref{amp:3}, which therefore (as explained in Remark \ref{rema:1}) leads to the particular representation \eqref{amp:7} of the Dirac measure.  This regularization is physically well motivated by the notion that a free particle corresponds to a `plane-wave' of infinite extent.

There is no solution to the second problem if the $\delta$-function in the amplitude \eqref{amp:6} is interpreted as a Schwartz distribution.  Consequently, the conventional solution to the third problem, i.e., the integration of the $\delta^2$-function in the differential cross-section \eqref{pro:6}, which is explained in Secs.\,\ref{dif:0} and \ref{jus:0}, and which is essentially equivalent to `Fermi's golden rule,' is an \emph{ad hoc} calculation unrelated to distribution theory whose physical justification is that it gives an answer that agrees with the data.\footnote{As remarked in \cite[p.\,101]{BJORK1964-}, the use of `wave-packets' to represent the incident and emerging particles implies that the appearance of squares of $\delta$-functions is avoided. (See also, \cite[p.\,105--106]{PESKI1995-}.)  
Mathematically, this is equivalent to chosing a smooth rather than sharp regularization of the Dirac $\delta$-function, and working on the `representatives' rather than on the `distribution' itself.  This bypasses problems 2 and 3, but does not solve them.}

A mathematically consistent solution to all three problems is to redefine the transition amplitude \eqref{amp:1}, and thus \eqref{amp:6}, as a nonlinear generalized function in a Colombeau algebra.  Instead of  \eqref{amp:6} we therefore write
\begin{equation}\label{dis:1}
     S_{fi} = 2\pi \frac{Ze^2}{V} \frac{1}{\sqrt{E_f E_i}}
                   \frac{1}{q^2} \rho_\epsilon(E_f-E_i),
\end{equation}
where $\rho_\epsilon(E) = \rho(E/\epsilon)/\epsilon$ is the scaled mollifier associated to the damper $\FOU{\rho}$ in $\eqref{rem:1}$ which replaces the sharp cut-off introduced in Remark \ref{rema:1}, and where $1/\epsilon$ corresponds to $T/2$ as remarked in Sec.\,\ref{rem:0}.2.  Therefore, the usual quantum mechanical rule \eqref{pro:1} leading from amplitudes to probabilities yields
\begin{equation}\label{dis:2}
   W_{fi} =   \lim_{T \rightarrow \infty} 
             8\pi \frac{Z^2\alpha^2}{E_iV} \frac{1}{q^4}
              \frac{d^3p_f}{E_f} |\rho_{2/T}|^2(E_f-E_i),
\end{equation}
where the limit $T \rightarrow \infty$,  is taken at the end as in the conventional calculation.

It remains then, just like in the conventional theory, to insure that the regularization implied by the damper $\FOU{\rho}$ yields the proper physical result, i.e., Rutherford's formula.  This has been done in Sec.\,\ref{rig:0} where, expressed as conditions on its inverse Fourier transform $\rho$, the conditions on $\FOU{\rho}$ are given by \eqref{rig:8}, i.e.,
\begin{gather}
 \nonumber
    \rho(z) \in \mathcal{S}(\mathbb{C}),
    \quad
    \int dz~\rho(z) = 1,
    \quad
    \int dz~z^n\rho(z) = 0,
    \quad \forall n=1,...,q,\\
\label{dis:3}
    \rho(0) = \frac{1}{\pi},
    \quad
    \int dz~ |\rho|^2(z)  = \rho(0).
\end{gather}
The conditions on the first line of this set are necessary in order that the square of the $\delta$-function whose representative is $\rho_{2/T}$ is well defined, so that it is possible to calculate the probability by squaring the amplitude.  These conditions are equivalent to the statement that $\rho_{2/T}$ is an element of a Colombeau algebra $\mathcal{G}$, i.e., a nonlinear generalized function. 

The conditions on the second line are restricting the  $\mathcal{G}$-function $\rho$, or equivalently its Fourier transforms $\FOU{\rho}$, to a particular subset of $\mathcal{G}$ such that \eqref{dis:2} yields the Rutherford formula.  In combination with the conditions on the first line, they insure that the regularization implied by $\FOU{\rho}$ corresponds to a special class of smooth wave-packets whose properties are equivalent to a plane-wave of infinite spatial and time extent.\footnote{In the present paper the focus is on the time-integral in Eq.\,\eqref{amp:3} because the spatial Fourier transform of the Coulomb potential, Eq.\,\eqref{amp:4}, is well defined.  In a general scattering problem, the integrations over all four dimensions are likely to need some kind of regularization.}
 
As was remarked is Sec.\,\ref{rem:0}.1, it is not easy to find a simple explicit example of a suitable mollifier $\rho$ satisfying all conditions \eqref{dis:3}, but there is not doubt that such mollifiers, and their corresponding dampers $\FOU{\rho}$, exist.  The physically interesting question which is not addressed in the present paper is whether this class of mollifier/dampers is just a subset of $\mathcal{G}$ such that first order time-dependent perturbation calculations make sense mathematically, or whether it corresponds to a physically significant redefinition of the notion of quantum-mechanical state with potentially far reaching implications for the foundations of quantum theory.


\newpage 

\section{References}
\label{bib:0}


\begin{enumerate}

\bibitem{DIRAC1927-} P.A.M. Dirac, \emph{The physical interpretation of quantum mechanics}, Proc. Roy. Soc. {\bf A 113} (1927) 621--641.

\bibitem{LANCZ1926A} C. Lanczos, \emph{On a field theoretical representation of the new quantum mechanics}, Zeits. f. Phys. {\bf 35} (1926) 812--830, translated and commented in W.R. Davis et al., Cornelius Lanczos Collected Published Papers with Commentaries, Vol.\,III (North Carolina State University, 1998) 2-858 -- 2-895.

\bibitem{COLOM1992-} J.F. Colombeau, Multiplication of Distributions, Lecture Notes in Mathematics {\bf 1532} (Springer Verlag, 1992) 184~pp.

\bibitem{GROSS2001-} M. Grosser, M. Kunzinger, M. Oberguggenberger, and R. Steinbauer, Geometric Theory of Generalized Functions with Applications to General Relativity, Mathematics and its Applications {\bf 537} (Kluwer Acad. Publ., Dordrecht-Boston-New York, 2001) 505~pp. 

\bibitem{GSPON2006B} A. Gsponer, \emph{A concise introduction to Colombeau generalized functions and their applications in classical electrodynamics}, Report ISRI-06-02 (2006) 19~pp. e-print arXiv:math-ph/0611069. 

\bibitem{GSPON2008A} A. Gsponer, \emph{The sequence of ideas in a re-discovery of the Colombeau algebras}, Report ISRI-08-01 (2008) 28~pp. e-print arXiv:0807.0529. 

\bibitem{BJORK1964-} J.D. Bjorken and S.D. Drell, Relativistic Quantum Mechanics (McGraw-Hill, New York, 1964) 299~pp.

\bibitem{SAKUR1994-} J.J. Sakurai, Modern Quantum Mechanics (Addison-Wesley, 1994) 500~pp.

\bibitem{JAUCH1955-} J.M. Jauch and F. Rohrlich, The Theory of Photons and Electrons (Addison-Wesley, New York, 1955) 489~pp.

\bibitem{GRADS1965-} I.S. Gradshteyn and I.M. Ryzhik, Table of Integrals, Series, and Products (Academic Press Inc, New York, 1965) 1086~pp.

\bibitem{SCHUC1991-} T. Sch\"ucker, Distributions, Fourier transforms, and Some of Their Applications to Physics (World Scientific, Singapore, 1991) 167~pp.

\bibitem{COLOM2008-} J.F. Colombeau and A. Gsponer, \emph{The Heisenberg-Pauli canonical formalism of quantum field theory in the rigorous setting of nonlinear generalized functions (Part I)} (2008) 108~pp.  e-print arXiv:0807.0289.

\bibitem{PESKI1995-} M.E. Peskin and D.V. Schroeder, An Introduction to Quantum Field Theory (Perseus Books, Reading Massachusetts, 1995) 842~pp.



\end{enumerate}
\end{document}